\documentclass[fleqn,usenatbib]{mnras}

\usepackage{newtxtext,newtxmath}
\usepackage[T1]{fontenc}

\DeclareRobustCommand{\VAN}[3]{#2}
\let\VANthebibliography\thebibliography
\def\thebibliography{\DeclareRobustCommand{\VAN}[3]{##3}\VANthebibliography}

\usepackage{graphicx}	
\usepackage{amsmath}
\newtheorem{theorem}{Theorem}

\newcommand{\eb}{\mathbfit{e}}

\newcommand{\rb}{\mathbfit{r}}

\newcommand{\Bb}{\mathbfit{B}}

\title[Maximising Open Magnetic Flux]{Maximising the Open Magnetic Flux of a Star}

\author[A. R. Yeates]{
Anthony R. Yeates,$^{1}$\thanks{E-mail: anthony.yeates@durham.ac.uk (ARY)}
\\
$^{1}$Department of Mathematical Sciences, Durham University, South Road, Durham DH1 3LE, UK
}

\date{Accepted XXX. Received YYY; in original form ZZZ}

\pubyear{\the\year{}}

\begin{document}
\label{firstpage}
\pagerange{\pageref{firstpage}--\pageref{lastpage}}
\maketitle

\begin{abstract}
We consider how to distribute the magnetic flux on a star's surface in order to maximise the proportion of open magnetic flux predicted by  the potential field source surface model. The proportion of open flux is shown to be maximised by concentrating all of the positive and negative flux at antipodal points. For a typical source surface height, this enhances the open flux by 35\% compared to a dipolar distribution. Even if the flux is spread over finite polar caps -- as on the Sun around Solar Minimum -- the open flux is still substantially enhanced compared to the dipole. Nevertheless, we show that the dipole approximation over-estimates the open flux for this kind of flux distribution.
\end{abstract}




\section{Introduction} \label{sec:intro}

The open magnetic flux of a star -- such as the Sun -- is the unsigned magnetic flux that escapes from its corona into interplanetary space. We denote it by
\begin{equation}
    \Phi_1 = r_1^2\oint_{S_1}\big|B_r(r_1,\theta,\phi)\big|\,d\Omega,
    \label{eq:osf}
\end{equation}
where $\Bb$ is the star's magnetic field and $S_1$ is a suitable spherical surface of integration whose radius $r_1$ is high enough above the stellar surface so that all magnetic field lines above it are pulled out radially into the stellar wind. As the source of the heliospheric magnetic field, $\Phi_1$ has an important influence on conditions around planets including Earth \citep[e.g.,][]{lockwood2009,owens2013}. In stellar astronomy, the open flux is also important for understanding long-term evolution because of angular momentum loss by magnetic braking \citep[e.g.,][]{schatzman1962,reville2015, see2018, pantolmos2020}.

The amount of open flux from a stellar corona depends both on the stellar surface flux distribution and on the magnetic structure in the corona itself, where the presence of electric currents can increase the amount of open versus closed flux \citep[e.g.,][]{wang1988,yeates2010}. However, in this paper we consider only the effect of the surface distribution, assuming the minimum-energy corona in the form of the potential field source surface model \citep{schatten1969,altschuler1969}. This approximation has frequently been used to model open flux of both the Sun \citep[e.g.,][]{wang1988,lee2011,arden2014,virtanen2020} and other stars \citep[e.g.,][]{see2018}. In many other studies, the potential field is used to initialise a more realistic magnetohydrodynamic model, but the magnetic topology is effectively inherited from the potential field \citep[e.g.,][]{reville2015,garraffo2016,mikic2018,perri2022}.

The seminal work on how surface flux distributions affect open flux on the Sun was by \citet{wang2002jgr}. In particular, these authors showed that -- at least in the potential model -- high-degree harmonics (small angular scales) in the magnetic field are irrelevant for the open flux, which in fact is well approximated at most times by the dipole strength alone \citep[see also][]{see2018, yoshida2023, tahtinen2024}. This is understood mathematically from the fact that the radial laplace eigenfunctions decay faster with height for higher degrees. Physically, the basic picture is that higher degree modes have more polarity reversals on the stellar surface, leading to magnetic loops that close down at lower heights.

This dominance of dipole strength is consistent with the finding of \citet{finley2018} that spin-down torque for most low-mass stars is dominated by the dipole component. It has also been used to explain why there was a large increase in open flux from the Sun during 2014 \citep{wang2015, yoshida2023, heinemann2024, tahtinen2026}. Several large bipolar active regions emerged at a single ``active longitude'' so that their dipole vectors added together and increased the Sun's overall dipole strength, whereas usually flux emergence is more random in longitude so that the stronger east-west dipole contributions tend to average out.

This short paper is motivated by a simple mathematical question: how should we distribute a fixed amount of (unsigned) flux on the stellar surface in order to maximize the open flux, at least in a potential field? Given how well the dipole strength approximates the open flux, one might suspect that a dipole would be the optimum distribution. However, we shall prove that this is not the case and that a star can do better by concentrating its positive and negative flux at opposing poles. After defining the potential field equations in Section \ref{sec:pfss}, this optimum result is given in Section \ref{sec:optimum}. Section \ref{sec:top-knot} then presents a family of more realistic magnetic fields where the flux is concentrated not at singular points but in finite-width ``polar caps''. Finally, Section \ref{sec:dip} notes an interesting point about the dipole approximation for $\Phi_1$ in such fields. Brief conclusions are given in Section \ref{sec:conclusion}.

\section{Coronal magnetic field model} \label{sec:pfss}

Throughout this paper we adopt the traditional ``potential field source surface'' model for the magnetic field in the stellar corona \citep{schatten1969,altschuler1969}, taken to be the spherical shell volume $r\in[r_0, r_1]$. This model uniquely defines the magnetic field $\Bb(r,\theta,\phi)$ through the following boundary value problem:
\begin{enumerate}
    \item $\boldsymbol{\nabla}\cdot\Bb=0$ and $\boldsymbol{\nabla}\times\Bb=\mathbfit{0}$ for $r\in(r_0, r_1)$;
    \item $B_\theta(r_1,\theta,\phi)=B_\phi(r_1,\theta,\phi)=0$ on the outer boundary;
    \item $B_r(r_0,\theta,\phi) = f(\theta,\phi)$ is prescribed on the stellar surface.
\end{enumerate}
The solution for $\Bb$ may be obtained by standard techniques as a spherical harmonic expansion \citep[e.g.,][]{mackay2012}. Whilst our optimisation will allow for general two-dimensional distributions $f(\theta,\phi)$, we shall require explicit expressions only for axisymmetric distributions where $\partial f/\partial\phi\equiv 0$. In that case $f$ can be expanded in Legendre polynomials as $f(\theta) = \sum_{\ell = 0}^\infty f_\ell P_\ell(\cos\theta)$. We will need the radial component of the solution, which is
\begin{align}
    &B_r(r,\theta) = \nonumber\\
    &\quad\sum_{\ell = 0}^\infty f_\ell\frac{\ell(r/r_0)^{\ell-1}(r_0/r_1)^{2\ell+1} + (\ell+1)(r_0/r)^{\ell+2}}{\ell + 1 + \ell(r_0/r_1)^{2\ell+1}}P_\ell(\cos\theta).
    \label{eq:br}
\end{align}
Using $\boldsymbol{\nabla}\cdot\Bb=0$ inside the star implies that we should only consider $f(\theta,\phi)$ distributions with vanishing signed integral over the sphere, in which case $f_0=0$ in the above expansions. However, the monopole term will be useful in intermediate calculations. Finally, we define the unsigned surface flux by
\begin{equation}
\Phi_0 = r_0^2\oint_{S_0}\big|f(\theta,\phi)\big|\,d\Omega = r_0^2\oint_{S_0}\big|B_r(r_0,\theta,\phi)\big|\,d\Omega,
\label{eq:phi0}
\end{equation}
where $d\Omega=\sin\theta\,d\theta d\phi$ and $S_0$ denotes the inner boundary $r=r_0$. This is analogous to the open flux $\Phi_1$ in Equation \eqref{eq:osf}, where the integral was over the outer boundary $S_1$ (the surface $r=r_1$). Our main objective is to maximise the ratio $\Phi_1/\Phi_0$ by choosing the appropriate boundary distribution $f(\theta,\phi)$.

\section{Optimum distribution} \label{sec:optimum}

The following result is proven in Appendix \ref{sec:app}.

\begin{theorem} \label{thm1}
    For any continuous surface distribution $f(\theta,\phi)$,
    $$
    \Phi_1/\Phi_0 \leq \sum_{\substack{\ell=1\\\textrm{$\ell$ odd}}}^\infty \frac{(2\ell+1)^2(r_0/r_1)^\ell}{\ell + 1 + \ell(r_0/r_1)^{2\ell+1}}\int_0^1P_\ell(x)\,dx,
    $$
    which is the ratio obtained for $$f(\theta) = B_0\big[\delta(\cos\theta-1) - \delta(\cos\theta+1)\big].$$
\end{theorem}


In other words, the proportion of open flux is maximised by concentrating all of the positive and negative flux at antipodal points on the sphere, with the limiting distribution consisting of a pair of Dirac delta functions. Note that the constant $B_0$ sets the amount of flux, but drops out of the ratio $\Phi_1/\Phi_0$. For computation, the remaining integral may be written \citep[e.g.,][p.306]{whittaker2020} in the closed form
\begin{equation}
     \int_0^1P_\ell(x)\,dx = \frac{(-1)^q(\ell-1)!}{4^q(\ell+1)(q!)^2}, \quad \textrm{where} \quad q=\frac{\ell-1}{2}.
    \label{eq:pl}
\end{equation}

\begin{figure}
    \centering
    \includegraphics[width=\linewidth]{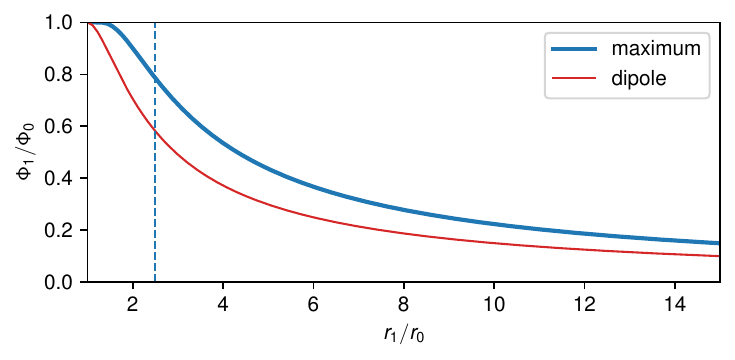}
    \caption{Dependence of the maximum ratio $\Phi_1/\Phi_0$ from Theorem \ref{thm1} on the domain height $r_1/r_0$ (with the sum computed up to $\ell=61$), as well as the corresponding ratio for the dipole distribution $f(\theta)=B_0\cos\theta$ (Section \ref{sec:top-knot}). The dashed line shows a typical source surface height of $r_1=2.5r_0$.}
    \label{fig:ratio}
\end{figure}

Figure \ref{fig:ratio} (thick blue curve) shows how the maximum ratio $\Phi_1/\Phi_0$ varies as a function of $r_1/r_0$. It decays to zero as $r_1/r_0\to\infty$, but the decay is not especially rapid. Moreover, for fixed $r_1/r_0$ it is substantially higher than a dipole solution with $f(\theta)=\cos\theta$ (thin red curve). For the typical source surface height of $r_1=2.5r_0$, we find $\Phi_1/\Phi_0=0.7855$ for the antipodal distribution (to four decimal places), compared to only $\Phi_1/\Phi_0=0.5814$ for the dipole -- a $35\%$ increase. In the next section, we consider a family of ``polar cap'' distributions between these two extremes.

\section{Finite polar caps} \label{sec:top-knot}

\begin{figure*}
    \centering
    \includegraphics[width=\linewidth]{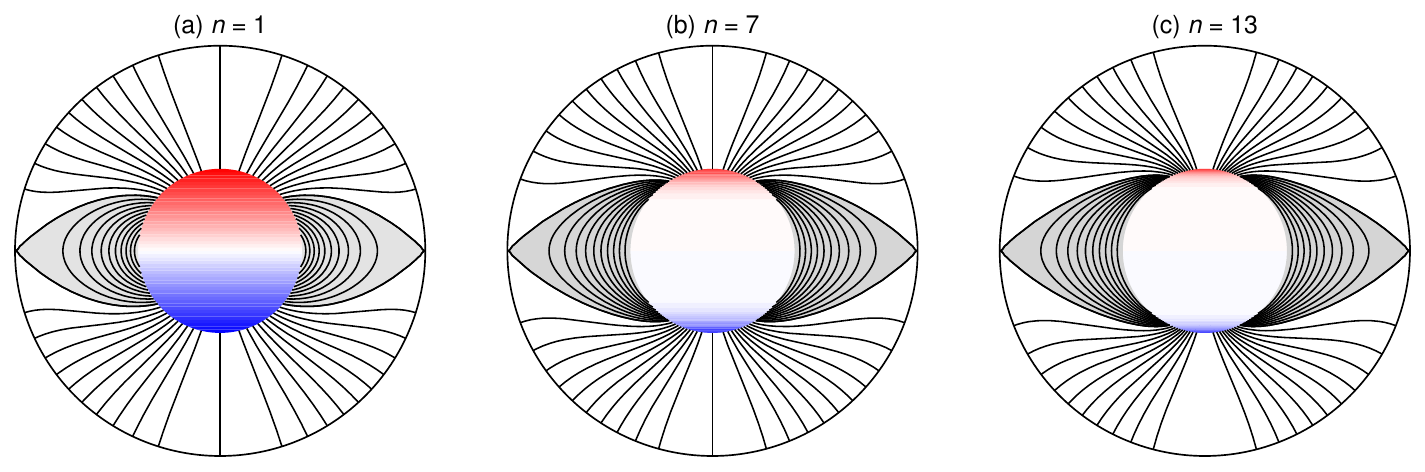}
    \caption{Polar cap solutions $f(\theta)=B_0\cos^n\theta$ for $n=1, 7$ and $13$. For these axisymmetric magnetic fields, field lines are plotted as contours of $r\sin\theta A(r,\theta)$, where $A$ is a flux function satisfying $\Bb=-\boldsymbol{\nabla}\times(A\eb_\phi)$. Red/blue shows positive/negative $f$ on the stellar surface and gray shading shows closed field.}
    \label{fig:fl}
\end{figure*}

Around Solar Minimum, the solar surface is often free of active regions and the large-scale surface magnetic flux is concentrated in polar caps of finite width, due to the poleward meridional flow \citep{sheeley1989, yeates2023}. We can model such a distribution by a simple function of the form
\begin{equation}
    f(\theta) = B_0\cos^n\theta, \quad \textrm{for $n=1,3,5,\ldots$}
    \label{eq:topknot}
\end{equation}
Three potential field solutions from this family are shown in Figure \ref{fig:fl}. The dipole corresponds to $n=1$, while increasing $n$ concentrates the flux more and more  at the poles. The limit $n\to\infty$ would be the antipodal distribution in Theorem \ref{thm1}.

\begin{figure}
    \centering
    \includegraphics[width=\linewidth]{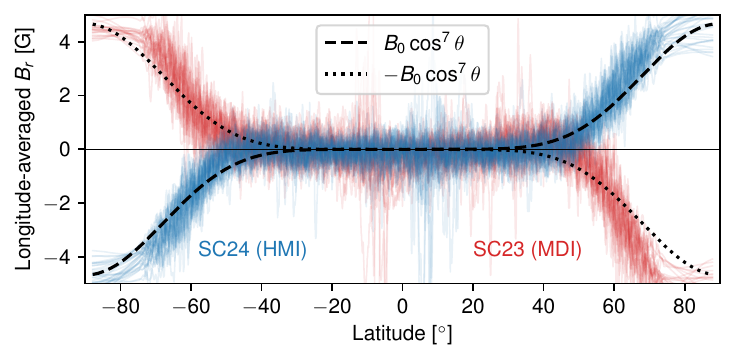}
    \caption{Latitudinal profile of longitude-averaged $B_r$ on the solar surface obtained from line-of-sight synoptic magnetograms with SOHO/MDI (red; Carrington rotations 2060 to 2084, August 2007 to May 2009) and SDO/HMI (blue; Carrington rotations 2201 to 2233, February 2018 to July 2020).}
    \label{fig:obspolar}
\end{figure}

The real Solar Minimum magnetic field is closest to $n=7$ \citep{svalgaard1978, wang1988}. This is illustrated by Figure \ref{fig:obspolar}, which overlays latitudinal profiles from a number of Carrington synoptic maps around the end of Solar Cycle 23 (in red) and the end of Solar Cycle 24 (in blue). Pole-filled maps were used (hence the smoother curves at high latitudes): from SOHO/MDI \citep{scherrer1995} for Solar Cycle 23 and SDO/HMI \citep{scherrer2012} for Solar Cycle 24.

\begin{figure}
    \centering
    \includegraphics[width=\linewidth]{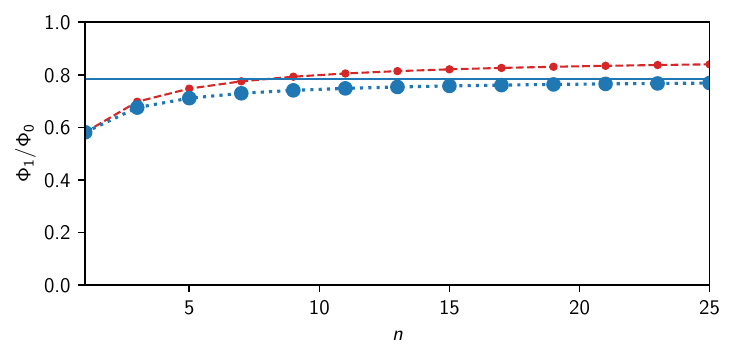}
    \caption{Dependence of $\Phi_1/\Phi_0$ on $n$ for the polar cap distributions $\cos^n\theta$ (large blue dots), fixing $r_1=2.5r_0$. The horizontal line shows the limiting ratio from Theorem \ref{thm1} while the small red dots show $\Phi_1/\Phi_0$ that would be estimated using only the dipole component of each surface distribution (Section \ref{sec:dip}).}
    \label{fig:caprat}
\end{figure}

We can find $\Phi_1/\Phi_0$ for the distributions of the form \eqref{eq:topknot} using the known expression of $\cos^n\theta$ \citep[][p.310]{whittaker2020} -- for odd $n$,
\begin{equation}
    \cos^n\theta = \sum_{\substack{\ell=1\\\textrm{$\ell$ odd}}}^n\frac{(2\ell+1)2^\ell n!\left[(n+\ell)/2\right]!}{\left[(n-\ell)/2\right]!(n+\ell+1)!}P_\ell(\cos\theta).
    \label{eq:cosn}
\end{equation}
Putting these coefficients as $f_{\ell}$ into \eqref{eq:br} and noting that $\Phi_0=4\pi r_0^2 B_0/(n+1)$ leads to
\begin{align}
    &\Phi_1/\Phi_0 = \nonumber\\
    &\sum_{\substack{\ell=1\\\textrm{$\ell$ odd}}}^n\frac{2^\ell (n+1)!\left[(n+\ell)/2\right]!(2\ell+1)^2(r_0/r_1)^\ell}{\left[(n-\ell)/2\right]!(n+\ell+1)!\big[ \ell + 1 + \ell(r_0/r_1)^{2\ell + 1}\big]}\int_0^1P_\ell(x)\,dx.
    \label{eq:caprat}
\end{align}
The behaviour of \eqref{eq:caprat} as a function of $n$ is illustrated in Figure \ref{fig:caprat}. One can show that this reduces to the expression in Theorem \ref{thm1} as $n\to\infty$ (see Appendix \ref{sec:app2}). At the opposite extreme, for a dipole ($n=1$), the sum in \eqref{eq:caprat} has only a single term, giving
\begin{equation}
    \Phi_1/\Phi_0 = \frac{3(r_1/r_0)^2}{2(r_1/r_0)^3 + 1},
    \label{eq:dip}
\end{equation}
in agreement with \citet{see2018}. For the solar-like case with $n=7$ and $r_1=2.5r_0$, we find $\Phi_1/\Phi_0=0.7296$, which is already a substantial increase compared to the dipole.

\section{Dipole approximation} \label{sec:dip}

As noted in Section \ref{sec:intro}, the open flux in the potential field (source surface) model is known to be well approximated by replacing $f(\theta,\phi)$ with only its dipole component, $f^{(1)}=\sum_{m=-1}^1f_{1m}Y_{1m}(\theta,\phi)$. Here we simply wish to point out a curiosity for the polar cap fields considered in this paper: for these fields, the dipole approximation over-estimates $\Phi_1$.

To see this, note \citep[cf.][]{wang1988} that the dipole component of the polar cap fields in \eqref{eq:topknot}, using \eqref{eq:cosn}, is 
\begin{equation}
    f^{(1)}(\theta) = \frac{3B_0}{n+2}\cos\theta,
\end{equation}
so the dipole approximation $\Phi_1^{(1)}$ of the open flux gives
\begin{equation}
    \Phi_1^{(1)}/\Phi_0 = \frac{3(n+1)}{2(n+2)}\left(\frac{3(r_1/r_0)^2}{2(r_1/r_0)^3 + 1}\right),
\end{equation}
This is shown (for $r_1=2.5r_0$) by the small red circles in Figure \ref{fig:caprat}. When the original field is a dipole, we have $n=1$ and the prefactor becomes unity, recovering \eqref{eq:dip}. But when the original field has $n>1$, the dipole approximation is seen to overestimate $\Phi_1/\Phi_0$ by 5--10\%. The estimate even exceeds the upper limit $0.7855$ when $n\geq 9$. The reason for the overestimate is that the terms in the sum \eqref{eq:caprat} alternate in sign due to the factor $(-1)^q$ in \eqref{eq:pl}. In the full sum, the $\ell=3$ term is negative and cancels part of the $\ell=1$ term, but this cancellation is missed if we account only for the dipole term. With more complex surface distributions, such perfect cancellation is unlikely, but it does explain why \citet{tahtinen2024} found $\Phi_1^{(1)}$ to overestimate $\Phi_1$ around Solar Minimum.

\section{Conclusions} \label{sec:conclusion}

We have shown that concentrating the surface magnetic flux into polar caps can substantially increase the open flux compared to a dipole, assuming the same potential field source surface model for the corona. This fundamental behaviour is expected to modulate the overall open flux even in a more energised (non-potential) corona. However, in a non-potential field -- where there are additional electric currents in the corona -- a star could, in principle, attain larger $\Phi_1/\Phi_0$, exceeding the maximum in Theorem \ref{thm1}. For example, a fully open ``split-monopole'' field (separated by a current sheet at the equator) would have $\Phi_1/\Phi_0 \approx 1$, but even this could be exceeded if U-shaped field lines were permitted that contributed to $\Phi_1$ but not $\Phi_0$.

For the Sun, quasi-static magneto-frictional modelling by \citet{yeates2010} to account for large-scale coronal currents suggests enhancement of $\Phi_1$ of 75--85\% compared to the potential field at Solar Maximum but only 25\% at Solar Minimum. And indeed such substantial enhancements are required to match near-Earth estimates of the open flux \citep{linker2017, arge2024}. However, more work is certainly required to constrain these non-potential models. More recent modelling suggests that $\Phi_1$ can also be enhanced by systematic small-scale vorticity on the solar surface \citep{mackay2022,klowss2025}, but ultimately fully time-dependent modelling of the heliospheric magnetic field may be required to properly account for the dynamical opening and closing of magnetic field \citep{lionello2026}.

\section*{Acknowledgements}

This work was supported by the Science and Technology Facilities Council [grant number UKRI1216]. Completion of the paper was stimulated by kind hospitality of Harvard-Smithsonian Center for Astrophysics through the ``van Ballegooijen lecture and residency'' in September 2026. SOHO is a project of international cooperation between ESA and NASA. HMI data are courtesy of NASA/SDO and the HMI science teams. The proof in Appendix \ref{sec:app3} was formulated with assistance from OpenAI's ChatGPT-5.6 Luna. I thank the anonymous reviewer for helpful feedback.

\section*{Data Availability}

The data used in Figure \ref{fig:obspolar} are publicly available from \url{http://jsoc.stanford.edu}.



\bibliographystyle{mnras}
\bibliography{yeates-osf}



\appendix

\section{Proof of Theorem 1} \label{sec:app}

\subsection{Ratio for the antipodal distribution} \label{sec:app1}

Firstly, we derive $\Phi_1/\Phi_0$ for the antipodal source distribution $f(\theta)=B_0\big[\delta(\cos\theta-1) - \delta(\cos\theta+1)\big]$ in Theorem \ref{thm1}. Integrating $|f|$ gives immediately $\Phi_0 = 4\pi r_0^2B_0$. To find $\Phi_1$, we use the standard Legendre expansion of the delta function \citep[][(1.17.22)]{NIST},
\begin{equation}
    \delta(\cos\theta-\cos\alpha) = \frac{1}{2}\sum_{l=0}^\infty(2l+1)P_l(\cos\theta)P_l(\cos\alpha),
    \label{eq:delta}
\end{equation}
to obtain
\begin{equation}
    f(\theta) = \frac{B_0}{2}\sum_{l=0}^\infty(2l+1)\big[1 - (-1)^\ell\big]P_l(\cos\theta).
\end{equation}
Reading off the coefficients $f_{\ell}$, the potential field solution \eqref{eq:br} is
\begin{equation}
    B_r(r_1,\theta) = B_0\sum_{\substack{\ell=1\\\textrm{$\ell$ odd}}}^\infty \frac{(2\ell+1)^2(r_0/r_1)^{\ell + 2}}{\ell + 1 + \ell(r_0/r_1)^{2\ell+1}}P_\ell(\cos\theta).
\end{equation}
Since $B_r$ is positive in the northern hemisphere and antisymmetric about the equator, the open flux must be
\begin{equation}
    \Phi_1 = 4\pi r_1^2B_0\sum_{\substack{\ell=1\\\textrm{$\ell$ odd}}}^\infty \frac{(2\ell+1)^2(r_0/r_1)^{\ell + 2}}{\ell + 1 + \ell(r_0/r_1)^{2\ell+1}}\int_0^1P_\ell(x)\,dx.
\end{equation}
Dividing by $\Phi_0$ gives the claimed expression
\begin{equation}
    \Phi_1/\Phi_0 = \sum_{\substack{\ell=1\\\textrm{$\ell$ odd}}}^\infty \frac{(2\ell+1)^2(r_0/r_1)^\ell}{\ell + 1 + \ell(r_0/r_1)^{2\ell+1}}\int_0^1P_\ell(x)\,dx.
    \label{eq:rat}
\end{equation}

\subsection{Kernel function} 

\begin{figure}
    \centering
    \includegraphics[width=\linewidth]{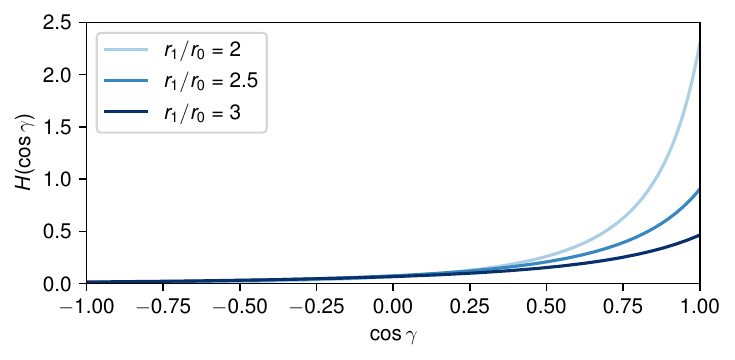}
    \caption{Dependence of $H(\theta,\phi;\theta',\phi')$ in \eqref{eq:greens} on the angle between $(\theta,\phi)$ and $(\theta',\phi')$, for three different domain heights.}
    \label{fig:greens}
\end{figure}

It will be helpful to introduce a kernel $H(\theta,\phi;\theta',\phi')$ such that
\begin{equation}
    B_r(r_1,\theta,\phi) = \oint_{S_0} H(\theta,\phi;\theta',\phi')f(\theta',\phi')\,d\Omega'.
    \label{eq:defkernel}
\end{equation}
This kernel encodes the contribution to $B_r$ at a point $(\theta,\phi)$ on the outer boundary from a particular point $(\theta',\phi')$ on the inner boundary. It depends only on the cosine of the angle between the two points, $\cos\gamma=\cos\theta\cos\theta' + \sin\theta\sin\theta'\cos(\phi-\phi')$. We will show that 
\begin{equation}
    H(\theta,\phi;\theta',\phi') = \frac{1}{4\pi}\sum_{\ell=0}^\infty\frac{(2\ell+1)^2(r_0/r_1)^{\ell + 2}}{\ell+1+\ell(r_0/r_1)^{2\ell+1}}P_\ell(\cos\gamma),
    \label{eq:greens}
\end{equation}
which is plotted as a function of $\cos\gamma$ in Figure \ref{fig:greens}. This kernel is non-negative and decays monotonically as the two points move apart on the sphere.

Note that $H$ is related to the standard Green's function $G(\rb,\rb')$ from potential theory \citep{jackson1975, sakurai1982}, which for $\rb = (r,\theta,\phi)$, $\rb' = (r_0,\theta',\phi')$, and our boundary conditions, would give
\begin{equation}
    \Bb(\rb) = -\boldsymbol{\nabla}\left(\frac{r_0^2}{4\pi}\oint_{S_0}G(\rb,\rb')f(\theta',\phi')\,d\Omega'\right). 
\end{equation}
The assumption of continuous $f$ in Theorem \ref{thm1} ensures that this solution exists \citep[cf.][Chap. IX]{kellogg1929}. Our kernel function in \eqref{eq:defkernel} is therefore related to $G$ by $H(\theta,\phi;\theta',\phi') = -r_0^2/(4\pi)\partial G/\partial r|_{r=r_1}$.

To derive \eqref{eq:greens}, note that $G$ lacks a simple closed formula in a spherical shell, so is usually expressed as a series expansion in spherical harmonics \citep{jackson1975}. Our boundary conditions (ii) and (iii) from Section \ref{sec:pfss} require Dirichlet conditions $G=0$ on $r=r_1$ and Neumann conditions $\partial G/\partial r=0$ on $r=r_0$. It is possible to obtain the resulting series for $G$ by adapting the argument in \citet[][Section 3.9]{jackson1975}, but one may obtain $H$ more easily just by considering a point source boundary distribution $f(\theta',\phi')$ in our known solution \eqref{eq:br}. Without loss of generality we can place this source at the north pole, so $f(\theta')=\delta(\cos\theta'-1)$. Then \eqref{eq:defkernel} gives
\begin{equation}
    H(\theta,\phi;0,0) = \frac{1}{2\pi}B_r(r_1,\theta),
\end{equation}
and from \eqref{eq:br} we know that
\begin{equation}
    B_r(r_1,\theta) = \frac{1}{2}\sum_{\ell=0}^\infty \frac{(2\ell+1)^2(r_0/r_1)^{\ell+2}}{\ell + 1 + \ell(r_0/r_1)^{2\ell+1}}P_\ell(\cos\theta),
\end{equation}
since $f_\ell = (2\ell+1)/2$ by \eqref{eq:delta}. By symmetry, moving the source to another location just changes $\theta$ to the new angle between $\rb$ and $\rb'$, so that $\cos\theta$ becomes $\cos\gamma$. Therefore the kernel must be \eqref{eq:greens}.

\subsection{Maximum ratio} \label{sec:app3}

To prove that \eqref{eq:rat} is the maximum possible ratio, consider a general inner boundary distribution $f(\theta,\phi)$ and split it into disjoint positive and negative parts:
\begin{equation}
    f(\theta,\phi) = f_+(\theta,\phi) - f_-(\theta,\phi),
\end{equation}
where $f_+,f_- \geq 0$ and $f_+f_- = 0$. A physically valid flux distribution must have no net flux, so that
\begin{equation}
    \oint_{S_0} f_+\,d\Omega = \oint_{S_0} f_-\,d\Omega = \frac{\Phi_0}{2r_0^2},
    \label{eq:compat2}
\end{equation}
with $\Phi_0$ as in \eqref{eq:phi0}. From \eqref{eq:defkernel}, with the shorthands $H'\equiv H(\theta,\phi;\theta',\phi')$ and $f_\pm'\equiv f_\pm(\theta',\phi')$, we have
\begin{equation}
    B_r(r_1,\theta,\phi) = \oint_{S_0}\big(H'f_+' - H'f_-'\big)\,d\Omega'.
\end{equation}
Using the properties of $f_+,f_-$ including \eqref{eq:compat2}, and defining the further shorthands $H''\equiv H(\theta,\phi;\theta'',\phi'')$ and $f_\pm''\equiv f_\pm(\theta'',\phi'')$, we can introduce a second integral by writing
\begin{align}
    B_r(r_1,\theta,\phi) &= \frac{2r_0^2}{\Phi_0}\left( \oint_{S_0}f''_-\,d\Omega''\right)\oint_{S_0}H'f'_+\,d\Omega' \nonumber\\
    &\qquad - \frac{2r_0^2}{\Phi_0}\left( \oint_{S_0}f''_+\,d\Omega''\right)\oint_{S_0}H'f'_-\,d\Omega'\\
    &= \frac{2r_0^2}{\Phi_0}\left( \oint_{S_0}f''_-\,d\Omega''\right)\oint_{S_0}H'f'_+\,d\Omega' \nonumber\\
    &\qquad - \frac{2r_0^2}{\Phi_0}\left( \oint_{S_0}f'_+\,d\Omega'\right)\oint_{S_0}H''f''_-\,d\Omega''\\
    &= \frac{2r_0^2}{\Phi_0}\oint_{S_0}\oint_{S_0}\big(H' - H''\big)f_+'f_-''\,d\Omega' d\Omega'',
\end{align}
so that
\begin{align}
    \Phi_1 &= \frac{2r_0^2r_1^2}{\Phi_0}\oint_{S_1}\left|\oint_{S_0}\oint_{S_0}\big(H' - H''\big)f_+'f_-''\,d\Omega'd\Omega'' \right|\,d\Omega\\
    &\leq \frac{2r_0^2r_1^2}{\Phi_0}\oint_{S_1}\oint_{S_0}\oint_{S_0}\big| H' - H''\big|f_+'f_-''\,d\Omega'd\Omega''d\Omega\\
    &=\frac{2r_0^2r_1^2}{\Phi_0}\oint_{S_0}\oint_{S_0} f_+'f_-''\left(\oint_{S_1}\big|H' - H''\big|\,d\Omega\right)\,d\Omega'd\Omega''\\
    &\leq \frac{\Phi_0}{2}\left(\frac{r_1}{r_0}\right)^2\sup_{(\theta',\phi'),(\theta'',\phi'')}\oint_{S_1}\big| H' - H''\big|\,d\Omega.
    \label{eq:osfbnd}
\end{align}
The last step used the fact that $f_+'f_-''\geq 0$ as well as \eqref{eq:compat2}.

Here $H'$ and $H''$ are the same kernel function \eqref{eq:greens}, but centred at two different points $(\theta',\phi')$ and $(\theta'',\phi'')$ on the sphere. As a function of $(\theta,\phi)$, each is symmetric about its own centre point (depending only on the angle $\gamma$). Moreover, the functional form is non-negative and decreasing with distance from the centre point, as shown in Figure \ref{fig:greens}. Therefore the integral of $|H'-H''|$ over $(\theta,\phi)$ is minimised when $(\theta',\phi')$ and $(\theta'',\phi'')$ are antipodal, because this minimises overlap of $H'$ and $H''$ and thus cancellation in the integral. Without loss of generality, we can evaluate the supremum by choosing $(\theta',\phi')$ and $(\theta'',\phi'')$ to be the north and south poles, whence substituting in from \eqref{eq:greens} gives
\begin{align}
    &\oint_{S_1}\big| H' - H''\big|\,d\Omega\nonumber\\
    &= \frac{1}{4\pi}\oint_{S_1}\left|\sum_{\ell=0}^\infty\frac{(2\ell+1)^2(r_0/r_1)^{\ell+2}}{\ell + 1 + \ell(r_0/r_1)^{2\ell+1}}\big[P_\ell(\cos\theta) - P_\ell(-\cos\theta)\big] \right|\,d\Omega\\
    &= 2\sum_{\substack{\ell=1\\\textrm{$\ell$ odd}}}^\infty \frac{(2\ell+1)^2(r_0/r_1)^{\ell + 2}}{\ell + 1 + \ell(r_0/r_1)^{2\ell+1}}\int_0^1P_\ell(x)\,dx.
\end{align}
Substituting in \eqref{eq:osfbnd} gives
\begin{equation}
    \Phi_1/\Phi_0 \leq \sum_{\substack{\ell=1\\\textrm{$\ell$ odd}}}^\infty \frac{(2\ell+1)^2(r_0/r_1)^{\ell}}{\ell + 1 + \ell(r_0/r_1)^{2\ell+1}}\int_0^1P_\ell(x)\,dx,
\end{equation}
which is precisely the ratio \eqref{eq:rat} obtained from the antipodal distribution. We conclude that this is an upper bound for the ratio, as claimed.

\section{Limiting expression} \label{sec:app2}

To obtain the large-$n$ limit of $\Phi_1/\Phi_0$ from equation \eqref{eq:caprat}, observe that
\begin{equation}
    \frac{[(n+\ell)/2]!}{[(n-\ell)/2]!} = \left(\frac{n-\ell+2}{2}\right)\left(\frac{n-\ell+4}{2}\right) \cdots \left(\frac{n+\ell-2}{2}\right)\left(\frac{n+\ell}{2}\right).
\end{equation}
This is a product of $\ell$ factors, so for large $n$ we have
\begin{equation}
    \frac{[(n+\ell)/2]!}{[(n-\ell)/2]!} \sim \frac{n^\ell}{2^\ell}.
\end{equation}
Similarly,
\begin{equation}
    \frac{(n+1)!}{(n+\ell+1)!} \sim \frac{1}{n^\ell}.
\end{equation}
Therefore the dependence on $n$ cancels from the coefficients of \eqref{eq:caprat} and the limit is given by the infinite sum in Theorem \ref{thm1}.


\bsp	
\label{lastpage}
\end{document}